\documentclass[10pt,twocolumn,letterpaper]{article}

\usepackage{cvpr} 
\usepackage{times}
\usepackage{epsfig}
\usepackage{graphicx}
\usepackage{amsmath}
\usepackage{amssymb}
\usepackage{booktabs} 
\usepackage{multirow} 
\usepackage{amssymb}
\usepackage{pifont}
\newcommand{\xmark}{\ding{55}}
\usepackage{makecell}
\usepackage{booktabs}
\usepackage{colortbl}
\usepackage{xcolor}
\usepackage{hyperref}

\begin{document}

\title{AirSignatureDB: Exploring In-Air Signature Biometrics in the Wild \\and its Privacy Concerns}

\author{
Marta Robledo-Moreno, Ruben Vera-Rodriguez, Ruben Tolosana,\\
Javier Ortega-Garcia, Andres Huergo, Julian Fierrez\\
\normalsize Biometrics and Data Pattern Analytics Lab, Universidad Autonoma de Madrid, Spain\\
{\tt\small \{marta.robledo, ruben.vera, ruben.tolosana, javier.ortega, julian.fierrez\}@uam.es}, \\
{\tt\small andres.huergo@estudiante.uam.es}
}

\maketitle
\thispagestyle{empty}

\begin{abstract}
   Behavioral biometrics based on smartphone motion sensors are growing in popularity for authentication purposes. In this study, AirSignatureDB is presented: a new publicly accessible dataset of in-air signatures collected from 108 participants under real-world conditions, using 83 different smartphone models across four sessions. This dataset includes genuine samples and skilled forgeries, enabling a comprehensive evaluation of system robustness against realistic attack scenarios. Traditional and deep learning-based methods for in-air signature verification are benchmarked, while analyzing the influence of sensor modality and enrollment strategies. Beyond verification, a first approach to reconstructing the three-dimensional trajectory of in-air signatures from inertial sensor data alone is introduced. Using on-line handwritten signatures as a reference, we demonstrate that the recovery of accurate trajectories is feasible, challenging the long-held assumption that in-air gestures are inherently traceless. Although this approach enables forensic traceability, it also raises critical questions about the privacy boundaries of behavioral biometrics. Our findings underscore the need for a reevaluation of the privacy assumptions surrounding inertial sensor data, as they can reveal user-specific information that had not previously been considered in the design of in-air signature systems.
\end{abstract}

\section{Introduction}
\label{sec:intro}

The field of Human-Computer Interaction (HCI) has recently moved toward more natural and intuitive authentication methods, taking advantage of the widespread presence of smartphones \cite{acien20sensors} and wearables \cite{marsico24sensors,sergio23food} in everyday life. 

Behavioral biometrics refer to identity traits derived from patterns in human actions \cite{straga22mobileB2C} rather than physical characteristics. Unlike physiological biometrics such as fingerprints or iris features, which tend to remain stable over time, behavioral traits are influenced by a range of factors that introduce high intra-user variability \cite{moises19survey,maiorana2025signature}. These include transient conditions (e.g., fatigue, emotional state, distraction), long-term changes (e.g., aging), and context shifts (e.g., posture, environment). Common behavioral biometrics include keystroke dynamics~\cite{KVC_benchmarkIEEEAcess23}, signature verification~\cite{tolo_deep}, gait and footstep recognition~\cite{delgado_gait, PAMIFootsteps2012} among others. Modern smartphones, equipped with accurate motion sensors, such as accelerometers and gyroscopes, have allowed the exploration of some of these characteristics for mobile user authentication \cite{stragapede_behavepass}. However, recent studies have also underscored the privacy risks associated with these sensor-based traits, highlighting important challenges in their deployment \cite{delgado_privacy}.

Among behavioral biometric traits, in-air signatures stand out as a user-friendly authentication method in which individuals perform a signature-like gesture in the air while holding a smartphone. These signatures are captured via the phone's built-in motion sensors, reflecting unique dynamic properties such as orientation or acceleration patterns. Other acquisition setups have used external sensors such as Kinect \cite{khoh_ihgs} or depth-based 3D hand tracking \cite{yushu_hand}, reinforcing the growing interest in contactless and privacy-aware biometrics.

In contrast to conventional signatures written on a surface, in-air gestures are inherently more dynamic and susceptible to alterations in fatigue, posture, or grip. Consequently, single-session studies frequently overestimate system performance by ignoring intra-user variability. To address this issue, data must be collected across multiple sessions arranged over days or weeks. This enables a robust evaluation under realistic temporal conditions. Furthermore, such evaluations must be based on data acquired \textit{in the wild}, using a variety of mobile devices and unconstrained environments, to fully capture the diversity and unpredictability of real-world usage. However, previous studies have often overlooked these crucial aspects, typically relying on controlled environments and single-session datasets. As a result, their reported performance tends to be overoptimistic and may not generalize well to real-world scenarios.

This work presents AirSignatureDB, a novel in-air signature dataset designed to overcome the limitations of prior studies and encourage progress in behavioral biometrics. Our main contributions are:

\begin{itemize}
    \item New in-air signature dataset with 108 users captured across four sessions under uncontrolled, real-world conditions using 83 different smartphone models, which is publicly available upon request. This allows for the study of intra-user variability, temporal drift, and hardware heterogeneity under realistic scenarios.

    \item We include both genuine samples and skilled forgeries for each user, enabling a comprehensive evaluation of system robustness to presentation attacks, an essential aspect of biometric security.

    \item We perform a baseline verification evaluation exploring both traditional and deep learning-based models for time series analysis.

    \item We demonstrate for the first time, to the best of our knowledge, the feasibility of reconstructing the 3D trajectory of in-air signatures from inertial data alone. This capability reduces the ambiguity associated with the inherently traceless nature of in-air gestures and offers potential value in legal or forensic contexts. However, it also raises important questions about the balance between biometric traceability and user privacy, which we address in the final discussion Section.
\end{itemize}

The remainder of the paper is structured as follows. Sec.~\ref{sec:database_description} provides an overview of the AirSignatureDB dataset, including acquisition protocol, device diversity, and forgery collection. Preprocessing techniques applied to inertial data are described in Sec.~\ref{sec:preprocessing}. Sec.~\ref{sec:verification_methods} presents the baseline models used for signature verification, including both classical and deep learning approaches. Sec.~\ref{sec: benchmark} details the experimental protocol and verification results. In Sec.~\ref{sec:privacy_discussion}, we explore the implications of 3D trajectory reconstruction on user privacy. Finally, Sec.~\ref{sec:conclusion} summarizes the findings and outlines directions for future work.

\section{Database Description}
\label{sec:database_description}

Modern smartphones are equipped with a variety of sensors, such as accelerometers, gyroscopes, and magnetometers. Considering that people carry these devices at all times, smartphones offer a powerful tool for behavioral biometric applications, including in-air signature verification.

\subsection{Existing Databases}
Table~\ref{tab:in_air_signature_comparisons} presents a comparison of the most relevant in-air signature datasets in the literature. Although we include the performance reported in the respective works, it is important to emphasize that these values are not directly comparable to ours. Each study employs different evaluation protocols, metrics, and conditions (ranging from session setup and data preprocessing to impostor generation and scoring methods). Most prior works, including Casanova~\textit{et al}.~\cite{casanova_realtime}, Bailador \textit{et al}.~\cite{bailador_patternrec}, Diep \textit{et al}.~\cite{diep_sigver3d}, and Yeo \textit{et al}.~\cite{yeo_realtime}, collected data in a single session under strictly controlled laboratory settings, using a fixed smartphone model. Consequently, their datasets do not account for temporal variability, contextual noise, or hardware heterogeneity, key factors in real-world biometric deployments.

\begin{table*}[h!]
    \centering
    \scriptsize
    \begin{tabular}{lccccccc}
    \hline
    & \textbf{\#Sessions} & \makecell{\textbf{Real-World} \\ \textbf{Variability}} & \textbf{\#Users} & \textbf{Capture Device} 
    & \makecell{\textbf{Trajectory} \\ \textbf{Reconstruction}} 
    & \makecell{\textbf{Public} \\ \textbf{Availability}} 
    & \makecell{\textbf{Reported} \\ \textbf{Performance}} \\ 
    \hline

    Casanova \textit{et al}. (2010)~\cite{casanova_realtime} & 1 & \xmark & 34 & iPhone 3G & \xmark & \xmark & 
    \makecell[l]{Random: Unknown \\ Skilled: EER = 2.5\%} \\
    \arrayrulecolor{black!30}\specialrule{0.1pt}{0pt}{0pt}

    Bailador \textit{et al}. (2011)~\cite{bailador_patternrec} & 1 & \xmark & 96 & iPhone 3G & \xmark & \xmark & 
    \makecell[l]{Random: EER = 2.12\% \\ Skilled: EER = 4.58\%} \\
    \arrayrulecolor{black!30}\specialrule{0.1pt}{0pt}{0pt}

    Casanova \textit{et al}. (2011)~\cite{guerra_intrauser} & 20 (over 4 months) & \xmark & 22 & iPhone & \xmark & \xmark & 
    \makecell[l]{Random: Unknown \\ Skilled: EER = 3.4\%} \\
    \arrayrulecolor{black!30}\specialrule{0.1pt}{0pt}{0pt}

    Diep \textit{et al}. (2015)~\cite{diep_sigver3d} & 1 & \xmark & 30 & Samsung S3 & \xmark & \xmark & 
    \makecell[l]{Random: Unknown \\ Skilled: EER = 0.8\%} \\
    \arrayrulecolor{black!30}\specialrule{0.1pt}{0pt}{0pt}

    Yeo \textit{et al}. (2015)~\cite{yeo_realtime} & 1 & \xmark & 50 & LG Nexus 5 & \xmark & \xmark & 
    \makecell[l]{Random: Unknown \\ Skilled: AER = 9.8\%} \\
    \arrayrulecolor{black!30}\specialrule{0.1pt}{0pt}{0pt}

    Li \textit{et al}. (2022)~\cite{gen_smartwatch} & 2 (1 week apart) & Partial & 22 & Smartwatch & \xmark & \xmark & 
    \makecell[l]{Random: Unknown \\ Skilled: EER = 0.83\%} \\
    \arrayrulecolor{black!30}\specialrule{0.1pt}{0pt}{0pt}

    Guo \textit{et al}. (2025)~\cite{guo_dataset} & 2 (1 week apart) & \xmark & 427 & Unspecified (only one) & \xmark & \checkmark & \makecell {Classification Acc \\ (Random) = 98\%} \\
    \arrayrulecolor{black!30}\specialrule{0.1pt}{0pt}{0pt}

    \textbf{Proposed Database} & 4 (at least 2 days apart) & \checkmark & 108 & 83 smartphone models & \checkmark & \checkmark & 
    \makecell[l]{Random: EER = 4.7\% \\ Skilled: EER = 2.3\%} \\
    \hline
    \end{tabular}
    \vspace{2mm}
    \caption{Comparison of in-air signature datasets in terms of acquisition protocol, real-world variability, trajectory reconstruction, public availability, and reported performance.  \textit{EER} denotes Equal Error Rate, \textit{AER} denotes Average Error Rate and \textit{Acc} stands for Accuracy.}
    \label{tab:in_air_signature_comparisons}
\end{table*}

Several prior works have attempted to incorporate temporal variation into their protocols, but with notable limitations. Li~\textit{et al}.\cite{gen_smartwatch} introduced a two-session setup spaced one week apart using a smartwatch, although the fixed device and controlled environment limited its realism. In \cite{guerra_intrauser}, Casanova \textit{et al}. collected a more extensive dataset over 20 sessions across four months with 22 users, explicitly addressing temporal variability; however, the data are not publicly available. More recently, Guo \textit{et al}.~\cite{guo_traceable} proposed a publicly available dataset focused on dominant hand detection. Although it spans two sessions and offers a novel take on traceability, the collection was still performed in non-real-world conditions using a single, unspecified device.

In contrast, AirSignatureDB is the only dataset that simultaneously fulfills several key criteria: it was collected under real-world conditions, involves 108 users, contains four sessions spaced, at least, two days apart, and uses 83 different \textit{Android} smartphone models. Furthermore, by including paired handwritten signatures, our dataset enables research in the reconstruction of 3D in-air trajectories aligned to ground-truth references. The dataset will be made publicly available, providing the biometric community with a realistic and reproducible benchmark for in-air signature verification.

\subsection{Acquisition Protocol}
To ensure continuous, real-world, large-scale data collection, we developed a custom \textit{Android} mobile app that remains publicly available and continues to accept contributions from new users\footnote{\url{https://play.google.com/store/apps/details?id=com.bida.behavepassdbuam}}. The application includes several biometric tasks, including both on-line handwritten and in-air signature capture. Data collection was in accordance with the ethical standards set out in the Declaration of Helsinki and was approved by the Ethics Committee of the Universidad Autónoma de Madrid (Approval No. CEI-142-3168).

A total of 198 participants completed the entire acquisition protocol and a detailed quality analysis of the recordings was performed to ensure data integrity and temporal consistency. This included filtering out sessions with sensor errors, incomplete recordings, or inconsistent signatures over time. After this post-processing stage, the final AirSignatureDB includes data from 108 users.  This database is publicly available upon request. The selected subset maintains demographic diversity: ages range from 15 to 65 years (mean = 30, std = 12.3), with 54 ``Male" (50\%), 52 ``Female" (48.2\%), and 2 ``Other” (1.8\%). In terms of hand preference, 98.2\% were right-handed and 1.8\% left-handed. 

The acquisition protocol involved two steps per session. First, users were asked to perform their handwritten signature twice on the touchscreen with the fingertip. Immediately after, they were instructed to replicate the same signature in the air, also twice, by holding the smartphone in one hand and performing the motion gesture freely in the space.

The dataset was collected over four sessions, with a minimum interval of two days between each. The actual time between sessions varied according to user availability, with an average of 4.12 days. This temporal spacing provides valuable information about intra-user variability, which is essential for evaluating long-term biometric consistency.

\subsection{Devices and Sensors}
\label{sec:devices}
The dataset includes samples from 83 different \textit{Android} smartphone models from 9 brands, with a clear predominance of \textit{Xiaomi} and \textit{Samsung}, which together account for more than 90\% of the devices used.

This diversity introduces natural variation in hardware characteristics such as sensor resolution, sampling rate, and internal filtering, which are factors rarely considered in controlled laboratory settings.  As a result, the dataset reflects the heterogeneity found in real-world mobile environments and provides a valuable testing environment for the development of universal biometric systems.

During the execution of tasks within the mobile application, data was recorded from multiple sources available on the device, including the touchscreen, accelerometer, linear accelerometer, gyroscope, magnetometer, Bluetooth status, GPU usage and gravity sensor. For the purpose of this study, the three sensors employed are: accelerometer, which measures 3-axis linear acceleration including gravity (\(m/s^2\)); linear accelerometer, which is a software-based sensor that applies a low-pass filter to the raw accelerometer to isolate gravity and extract only user-induced motion; and gyroscope, which captures 3-axis angular velocity (\(rad/s\)), providing information about device rotation during the signature. This combination enables a complete capture of both linear and rotational dynamics, offering a rich representation of the signing behavior.

\subsection{Skilled Forgeries}
\label{sec:skilled}
To assess the robustness of the system against imitation attacks, we designed a protocol considering skilled forgeries. Two trained impostors were tasked with imitating both the handwritten and in-air signatures of all users.

The imitation process was carried out in two phases: first, the impostors were shown a static image of the user's handwritten signature and asked to reproduce it twice. Then, they were provided with a real-time video of the handwritten signature execution, capturing motion characteristics such as stroke order and velocity. This two-step procedure ensured that forgeries captured not just the visual shape but also dynamic aspects of the signature.

\begin{figure}[tb]
  \centering
   \includegraphics[width=0.99\linewidth]{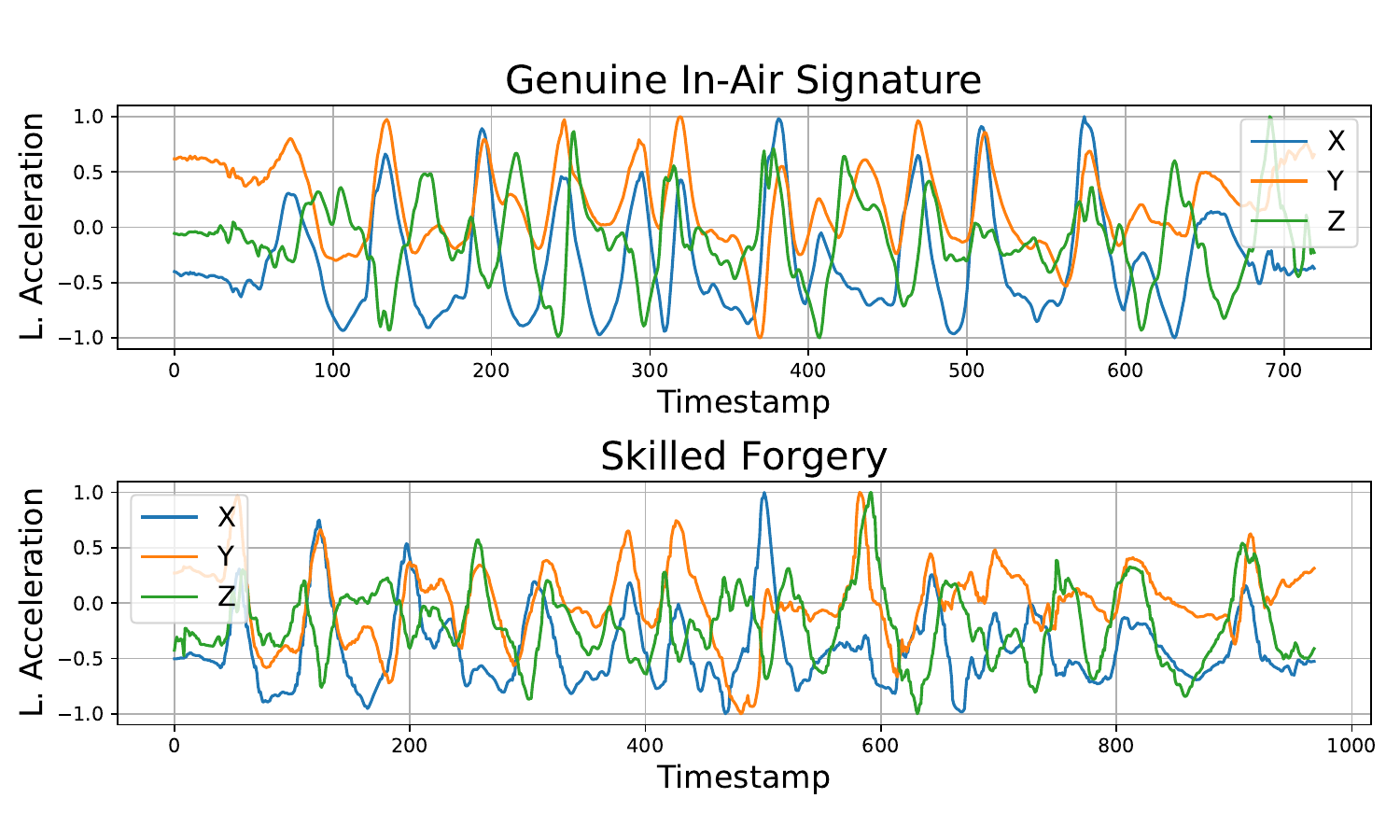}
       \caption{Linear acceleration signals (X, Y, Z) for a genuine in-air signature (top) and a skilled forgery (bottom) from the same user.}
   \label{fig:forgery_example}
\end{figure}

Fig.~\ref{fig:forgery_example} illustrates the genuine inertial signals from a user (top) and the corresponding forgery (bottom). The plots display linear acceleration in the three axes (X, Y, and Z) over time. Although accurately reproducing an in-air signature based only on its handwritten counterpart is very challenging, the imitation achieves a remarkably similar global motion pattern. Some differences in timing and acceleration profiles remain, but the overall structure is well preserved.

\section{Inertial Data Preprocessing}
\label{sec:preprocessing}

Raw motion signals captured from smartphone sensors are often affected by device variability. To ensure a reliable comparison of in-air signatures, a pre-processing pipeline is applied to standardize, clean, and improve input data before feature extraction \cite{moises19survey,maiorana2025signature}. This Section details the preprocessing steps employed in our system, including resampling to a fixed frequency, signal normalization, smoothing and applying a Movement Activity Detection (MAD) method in order to isolate informative gesture segments. 

\subsection{Sampling Frequency Variability}

An important feature of motion data acquisition on smartphones is the variability in the effective sampling rate of sensors. In the development of our \textit{Android} application, we explicitly requested the highest available sampling rate for each sensor to maximize temporal resolution. However, the actual frequency at which sensor events are delivered depends on various system-level factors, such as device manufacturer, operating system optimizations, background activity, and CPU load during acquisition.

This variability can affect the similarity and alignment of the signal. To ensure consistent signal representation across users, devices, and sessions, we resampled all sensor signals to a common frequency of 100 Hz. This value was selected based on the observed signals: it is high enough to preserve fine-grained signature dynamics while avoiding excessive interpolation in lower-rate devices.

\subsection{Normalization and Smoothing}
Each axis of the input signals (X, Y, Z) is first normalized using the Z-score normalization, which transforms data so that it has a mean of 0 and a standard deviation of 1. 

The signals are then smoothed using a moving average with a fixed window size of 5 samples. This simple smoothing step effectively reduces high-frequency noise while preserving the main motion patterns of the signature gesture.

\subsection{Movement Activity Detection}

Behavioral biometric signals captured under real-world conditions often contain segments of inactivity, noise, or unintentional motion that can degrade performance.

Our approach analyzes the energy of the signal over time using a sliding window. Given a motion signal $m(t)$ from the linear accelerometer, we compute the signal energy within overlapping windows of fixed duration. This sensor was chosen because it filters out the influence of gravity, preserving only the dynamic, user-induced motion. The global mean energy $\bar{E}$ across all windows is then used to determine an activity threshold. A window is labeled as containing signature motion if its energy exceeds $\tau \bar{E}$, with $\tau$ set empirically at 0.225 based on visual inspection.
\section{In-Air Signature Verification}
\label{sec:verification_methods}

Most methods for in-air signature verification are adaptations of techniques developed for 2D handwriting signature biometrics \cite{dtw_biom}. The evolution of these 2D signature methods can be traced back to the early 2000s, when core technologies such as DTW and HMMs were prominent, and benchmarks/competitions were established at major scientific events like ICB/IJCB \cite{svc2004} and ICDAR \cite{icdar2009}. More recent benchmarks, such as SVC-onGoing \cite{tolo22svc}, have seen dominant approaches based on deep learning \cite{tolo_deep, tolo21aaai} and transformers \cite{paula24swipe-trans}.

Previous works have explored different strategies for in-air signature verification. Diep \textit{et al}.~\cite{diep_sigver3d} used a global feature extraction approach where statistical features are computed from motion sensor data and classified using machine learning techniques. Benegui \textit{et al.}~\cite{benegui_cnn} proposed a system based on the application of a CNN to the raw motion sensor signals. Guo \textit{et al}. \cite{guo_dataset} explored three deep learning-based approaches: \textit{Fully Convolutional Network} (FCN), \textit{ResNet}, and \textit{InceptionTime}, which have demonstrated strong performance on time series classification tasks, including widely used UCR/EUA benchmarks \cite{DAU_UCR}.

In this work, we first evaluate the performance of the well-known \textit{Dynamic Time Warping} (DTW) algorithm for in-air signature verification, following the methodology adopted in previous studies~\cite{bailador_patternrec, casanova_realtime, yeo_realtime}. We also include a recurrent neural network model, motivated by its application in previous signature verification studies \cite{tolo_deep}. 

In addition, inspired by \cite{guo_dataset}, we investigate the same three deep learning-based approaches (FCN, ResNet and InceptionTime). Each neural network model is explored under multiple architectural and training configurations in order to identify the best-performing setup for our verification task. The ultimate goal is to use these models to learn compact embedding representations of signature dynamics. These representations are compared using cosine distance in a Siamese-style verification framework, where the goal is to minimize the distance between embeddings of genuine pairs and maximize it for impostor pairs. To ensure uniform input length across the network architectures, all input signals are either zero-padded or truncated to 1,000 samples, based on a preliminary analysis of the signature duration distribution in the dataset. Training is carried out using a Siamese network setup, where two parallel branches with shared weights process a pair of signature signals. The network is optimized using a contrastive loss function that operates on the cosine similarity between the resulting embeddings. Specifically, the loss encourages low cosine distance (i.e., high similarity) for genuine pairs and penalizes impostor pairs whose distance falls below a predefined margin.

\subsection{Dynamic Time Warping}

Dynamic Time Warping (DTW) is a widely used technique for time-series analysis, \textit{e.g.}, in signature biometrics~\cite{dtw_biom}. In our work, the final similarity score is computed using:

\begin{equation}
\text{score} = 1 - e^{-d/K}
\end{equation}
where $d$ is the DTW distance and  $K$ is the alignment path length.

\subsection{Recurrent Neural Network}

Recurrent Neural Networks (RNNs) are a class of models designed to process sequential data by maintaining temporal context across time steps.

The implemented RNN architecture consists of a two-layer bidirectional LSTM encoder, which processes the input sequence and captures both forward and backward temporal dynamics. The final hidden states from both directions are concatenated and projected into a 32-dimensional embedding space using a fully connected layer. During verification, the embeddings of both signatures are concatenated, forming a joint representation that is fed into a fully connected layer to produce a similarity score.

\subsection{Fully Convolutional Network}
FCN was originally developed for image segmentation tasks~\cite{long_FCN}, but was later effectively adapted for time series classification~\cite{DAU_UCR}. In this work, the implemented FCN architecture comprises three sequential 1D convolutional layers with kernel sizes of 8, 5, and 3, and output channels of 128, 256, and 128, respectively. Each convolution is followed by batch normalization and a ReLU activation. The final convolutional output is aggregated using an adaptive global average pooling operation, reducing the temporal dimension to one. A fully connected linear layer then projects the resulting vector to a 128-dimensional embedding space. The translational invariance of the system provides robustness to slight variations in gesture execution and sensor placement, which are common in real-world usage.

\subsection{Residual Network}

Residual Networks (ResNet) are a deep learning architecture designed to address the training difficulties of very deep models through the introduction of identity-based shortcut connections. Our ResNet-based architecture consists of three residual blocks, each containing three 1D convolutional layers with kernel sizes of 8, 5, and 3, and output channels of 64, 128, and 256, respectively. Each convolutional layer is followed by batch normalization and a ReLU activation, and shortcut connections are used to enable residual learning across mismatched dimensions. Unlike the sequential structure of the FCN, these residual connections help mitigate gradient vanishing and facilitate the training of deeper models. After the final block, the temporal dimension is reduced using adaptive global average pooling and the resulting feature vector is projected into a 128-dimensional embedding space via a linear layer. This architecture enables the network to preserve low-level features across layers while progressively learning abstract temporal patterns relevant to user identity.

\subsection{InceptionTime}
InceptionTime is a deep learning model specifically designed for time-series classification, inspired by the Inception architecture originally proposed for computer vision~\cite{fawaz_inceptiontime}. The architecture used in this study is made up of two stacked Inception blocks, each containing three sequential Inception modules. Each module applies a bottleneck layer followed by five parallel convolutional paths with kernel sizes of 3, 5, 8, 11, and 17, along with a max pooling branch, enabling the extraction of multi-scale temporal features. The outputs of all paths are concatenated, batch-normalized, and passed through ReLU activations. Residual connections join each block to improve gradient flow and stability. The final representation is obtained through adaptive global average pooling and projected into a 128-dimensional embedding space using a linear layer.
\section{AirSignatureDB Benchmark}
\label{sec: benchmark}
This Section describes the experimental setup used to benchmark the verification performance of in-air signatures under realistic conditions. We detail the data partitioning strategy, enrollment and evaluation protocol, attack scenarios considered, and different sensor configurations and models evaluated, as well as the results obtained.

\subsection{Experimental Protocol}
In order to rigorously evaluate the verification performance of the proposed dataset, an experimental protocol grounded in common biometric evaluation practices is defined. Out of the 108 users, 80\% were randomly selected for training and validation, while the remaining 20\% were reserved for testing. Within the training subset, we further apply an internal 80/20 split to construct validation folds for model selection and hyperparameter tuning.

Each user participated in four acquisition sessions. However, the first session is excluded from this benchmark due to frequent inconsistencies observed in the data, likely caused by user misunderstanding in the unsupervised, real-world capture setup. As a result, only data from sessions 2, 3, and 4 are used. Genuine signatures from sessions 2 and 3 are used for enrollment, and those from session 4 for verification. In the 1vs1 setup, one enrollment signature is matched with one test signature; in the 4vs1 setup, each test signature is compared against four enrollment samples, and scores are averaged.

Deep learning-based models are trained using both random and skilled impostor samples grouped as negative class instances. Random impostors are genuine samples from other users in the dataset, and skilled forgeries are created as described in Section \ref{sec:skilled}. This setup reflects a realistic deployment scenario in which the system must learn to reject any unauthorized input. However, an evaluation is performed separately for skilled and random impostor scenarios to better understand the behavior of the system under different threats.

\vspace{1em}

\subsection{Experimental Results}
The results presented in Table~\ref{tab:eer_transposed} are reported in terms of the Equal Error Rate (EER). They provide a comparison of the performance of in-air signature verification across various models, sensor modalities, and impostor types.

As observed in previous studies on signature verification~\cite{tolo_deep}, DTW usually outperforms deep learning in random-user comparison settings, probably because it directly exploits the temporal alignment between sequences.  A similar pattern emerges in the in-air signature scenario, where DTW achieves the best results in both the 1vs1 and 4vs1 random-user settings. In the 1vs1 case, the lowest EER is 9.0\%, while in the 4vs1 setup the EER drops to 4.7\% using only the linear accelerometer. This represents a relative improvement of 48\% EER, highlighting the benefit of additional enrollment data even in random-user scenarios.

In the skilled forgery scenario, the best results in the 1vs1 setting are achieved by both FCN with the accelerometer and linear accelerometer (8.9\% EER) and ResNet with the gyroscope (also 8.9\% EER). These models outperform others in this challenging setting when only one enrollment sample is available. In the 4vs1 setting, performance improves substantially: the best result is obtained with ResNet using sensor fusion (accelerometer, linear accelerometer, and gyroscope), reaching a remarkably low EER of 2.3\%. 

These results suggest that the gyroscope provides particularly discriminative behavioral information, likely capturing subtle patterns of device orientation and handling that are difficult to imitate. This is especially critical in the skilled impostor setting, where the trajectory of the signature may be mimicked, but the implicit dynamics, such as wrist rotation and grip, remain unique to the genuine user. The inability of attackers to replicate this information results in significantly improved performance.

Despite testing several configurations, the RNN models yielded consistently inferior results compared to convolutional architectures and the Inception-based one. The best configuration reached an EER of 11.3\%, significantly higher than the performance observed with ResNet. This suggests that, in their current form, RNNs may struggle to capture the fine-grained temporal dependencies relevant for in-air signature verification. Future work could explore the integration of preprocessing steps, such as temporal alignment, to enhance the effectiveness of sequential models.

It is also important to contextualize these results with respect to previous works in the literature. Although the best-performing deep learning models in this study achieve promising results, they do not outperform the lower error rates reported in more constrained datasets. This performance gap is expected given the challenging nature of the proposed dataset: it was collected in uncontrolled, real-world conditions, across multiple sessions, and using a heterogeneous pool of 83 smartphone models. These factors introduce significant variability in user behavior, sensor noise, and device-specific signal characteristics.

\begin{table*}[h]
\centering
\small
\begin{tabular}{llcccccccccc}
\toprule
\textbf{Forgery} & \textbf{Sensor} & \multicolumn{2}{c}{\textbf{DTW}} & \multicolumn{2}{c}{\textbf{FCN}} & \multicolumn{2}{c}{\textbf{ResNet}} & \multicolumn{2}{c}{\textbf{InceptionTime}} & \multicolumn{2}{c}{\textbf{RNN}} \\
             &                 & 1vs1 & 4vs1 & 1vs1 & 4vs1 & 1vs1 & 4vs1 & 1vs1 & 4vs1 & 1vs1 & 4vs1 \\
\cmidrule(lr){3-4} \cmidrule(lr){5-6} \cmidrule(lr){7-8} \cmidrule(lr){9-10} \cmidrule(lr){11-12}
\multirow{4}{*}{Random} 
  & Accelerometer        & 10.8 & 5.7  & 17.2 & 9.5  & 18.8 & 10.0  & 18.4 & 12.0 & 36.2 & 35.7 \\
  & Linear Accelerometer & \textbf{9.0}  & \textbf{4.7}  & 17.3 & 14.2 & \textbf{12.2} & 9.8  & 18.8 & 18.7 & \textbf{28.6} & \textbf{29.6} \\
  & Gyroscope            & 9.9  & 6.6  & \textbf{13.7} & 9.9  & 16.7 & 10.1 & \textbf{13.2} & \textbf{9.5} & 37.3 & 33.5 \\
  & Acc + L. Acc + Gyro  & 9.9   & 5.7  & 18.2 & \textbf{8.9} & 16.1 & \textbf{9.7} & 18.9 & 12.5 & 31.4 & 31.1 \\
\midrule
\multirow{4}{*}{Skilled} 
  & Accelerometer        & \textbf{19.3} & 14.4 & \textbf{8.9} & 5.4 & 12.2 & 7.4 & 16.6 & 12.2 & 20.3 & 28.4 \\
  & Linear Accelerometer & 22.4 & \textbf{12.5} & \textbf{8.9} & 7.4 & 9.8 & 3.2 & \textbf{9.8} & 7.1 & \textbf{14.5} & \textbf{11.3} \\
  & Gyroscope            & 20.5 & 20.8 & 9.5 & 5.3 & \textbf{8.9} & 5.0 & 10.4 & \textbf{4.4} & 28.5 & 23.8 \\
  & Acc + L. Acc + Gyro  & 20.8 & 16.0 & 11.9 & \textbf{3.8} & 9.5 & \textbf{2.3} & 12.5 & 6.8 & 16.3 & 22.0 \\
\bottomrule
\end{tabular}
\vspace{2mm}
\caption{Equal Error Rate (\%) for random and skilled impostors using DTW and deep learning models in 1vs1 and 4vs1 scenarios.}
\label{tab:eer_transposed}
\end{table*}

\section{Discussion: Trajectory Reconstruction and Privacy Implications}
\label{sec:privacy_discussion}

In-air signatures have frequently been regarded as a low-risk, privacy-friendly biometric modality due to the absence of a physical trace or visual representation. However, the findings of this study suggest that this assumption may no longer be valid. 

\subsection{Reconstruction of In-Air Signatures}
\label{sec:trajectory-reconstruction}
The reconstruction of 3D trajectories of in-air signatures has historically been considered an unsolved problem, not only due to the absence of ground-truth references, but also due to the intrinsic difficulty of the task. Previous works have attempted to reconstruct such 3D trajectories synthetically imitating the kind data coming from sensors such as Leap motion or Intel’s creative senz3D depth camera \cite{moises22-3d}, but no previous work has directly reconstructed 3D trajectories directly referring to smartphone sensors like here. A recent study by Guo et al. \cite{guo_traceable} introduced the concept of \textit{3D signature restoration}, which aims to generate traceable representations of in-air gestures. However, their method avoids performing double integration of the accelerometer signal, which is an essential step in recovering the trajectories from the acceleration data. Consequently, this approach does not generate accurate 3D reconstructions.

\begin{figure}[tb]
    \centering
    \includegraphics[width=\linewidth]{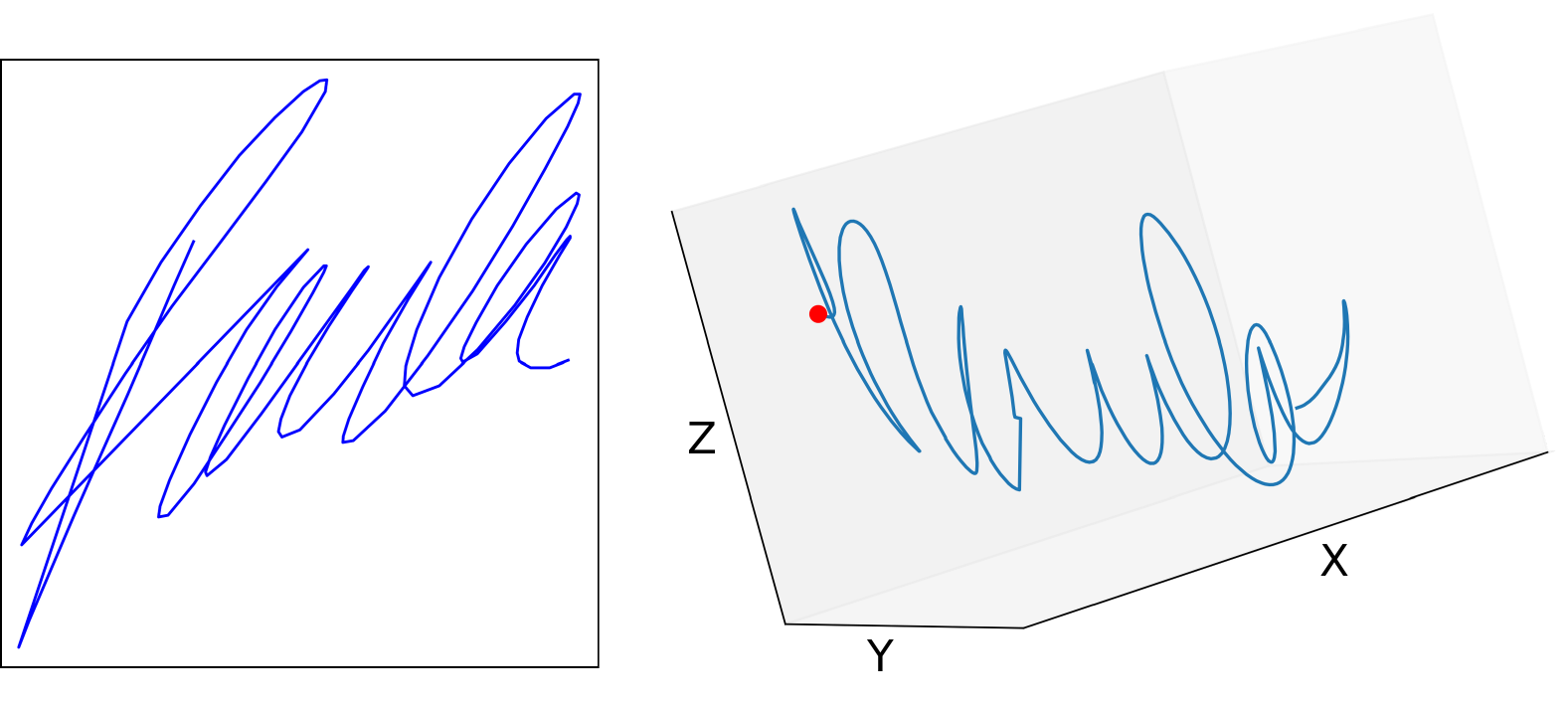}
    \caption{Example of 3D trajectory reconstruction from our dataset. The left panel shows the reference 2D handwritten signature, while the right panel shows the reconstructed in-air trajectory obtained from inertial sensor data. The red dot marks the starting point of the reconstructed gesture.}

    \label{fig:firmapaiula}
\end{figure}

In contrast to other in-air signature datasets, our approach incorporates paired in-air inertial data and handwritten two-dimensional signatures, enabling the capability of using these data as ground truth. Figure \ref{fig:firmapaiula} illustrates a reconstruction example from our dataset, where the corresponding handwritten signature serves as a reference to evaluate the precision of the reconstruction.

\subsubsection{Position Estimation from Inertial Sensors}
\label{sec:method}

The proposed approach involves the estimation of the position of a mobile phone in free space using solely the data provided by its built-in inertial sensors. To this end, a procedure was developed that utilizes only accelerometer and gyroscope data, avoiding the use of magnetometer data, which are often unreliable in indoor environments. 

In theoretical terms, the position can be obtained by performing a double integration of the acceleration over time:
\begin{equation}
\vec{v}(t) = \vec{v}(t_0) + \int_{t_0}^{t} \vec{a}(\tau) \, d\tau
\label{eq:velocity}
\end{equation}

\begin{equation}
\vec{r}(t) = \vec{r}(t_0) + \int_{t_0}^{t} \vec{v}(\tau) \, d\tau
\label{eq:position}
\end{equation}

\noindent
where \( \vec{a}(t) \) is the acceleration vector, \( \vec{v}(t) \) is the velocity vector and \( \vec{r}(t) \) is the position vector at time \( t \). The terms \( \vec{v}(t_0) \) and \( \vec{r}(t_0) \) denote the initial velocity and position, respectively, and \( t_0 \) is the initial time. These expressions assume that the acceleration is defined in a consistent, inertial frame of reference, typically the Earth reference frame.

In practice, however, the acceleration measured by the smartphone, \( \vec{a}_{\text{sensor}} \), is expressed in the local frame of the device (which is body-fixed). To transform this measurement into the inertial frame required for integration, it is necessary to estimate the orientation of the device over time. For this purpose, we used the angular velocity measured by the gyroscope, which describes how the device rotates in its local frame. The time evolution of the attitude quaternion \( \mathbf{q}(t) \) follows the differential equation:

\begin{equation}
\dot{\mathbf{q}}(t) = \frac{1}{2} \, \mathbf{q}(t) \otimes \begin{bmatrix} 0 \\ \vec{\omega}(t) \end{bmatrix}
\label{eq:quat_dynamics}
\end{equation}

\noindent
where \( \vec{\omega}(t) \) is the vector of angular velocity and \( \otimes \) denotes the multiplication of quaternions. Integrating this equation yields the device’s orientation over time, but the result is susceptible to drift due to sensor bias and noise.

To mitigate this issue, we applied the Madgwick filter \cite{hasan_madgwick}, which fuses gyroscope and accelerometer data to provide a drift-corrected estimate of the attitude. The gyroscope ensures accurate short-term tracking of orientation changes, while the accelerometer provides a long-term reference by sensing the direction of gravity. The output is a quaternion \( \mathbf{q}(t) \) that represents the device's orientation in the global frame. This allows the measured acceleration to be transformed into the inertial frame:

\begin{equation}
\vec{a}_{\text{global}}(t) = R(\mathbf{q}(t)) \cdot \vec{a}_{\text{sensor}}(t)
\label{eq:acc_global}
\end{equation}

\noindent
where \( R(\mathbf{q}(t)) \) is the rotation matrix associated with the quaternion \( \mathbf{q}(t) \). Once the acceleration is expressed in the Earth reference frame and corrected for gravity, velocity and position can be obtained through numerical integration. 

In order to accurately reconstruct the in-air signature, we first apply a first-order Butterworth high-pass filter to both the velocity and position signals. The cutoff frequency is adjusted per user and per signature, and is determined based on a frequency-domain analysis of the raw motion signal. A Fast Fourier Transform (FFT) is used to identify the dominant frequency components of the velocity and position, ensuring that the filter preserves the relevant signature dynamics while attenuating low-frequency drift.

\begin{figure}[tb]
    \centering
    \includegraphics[width=\linewidth]{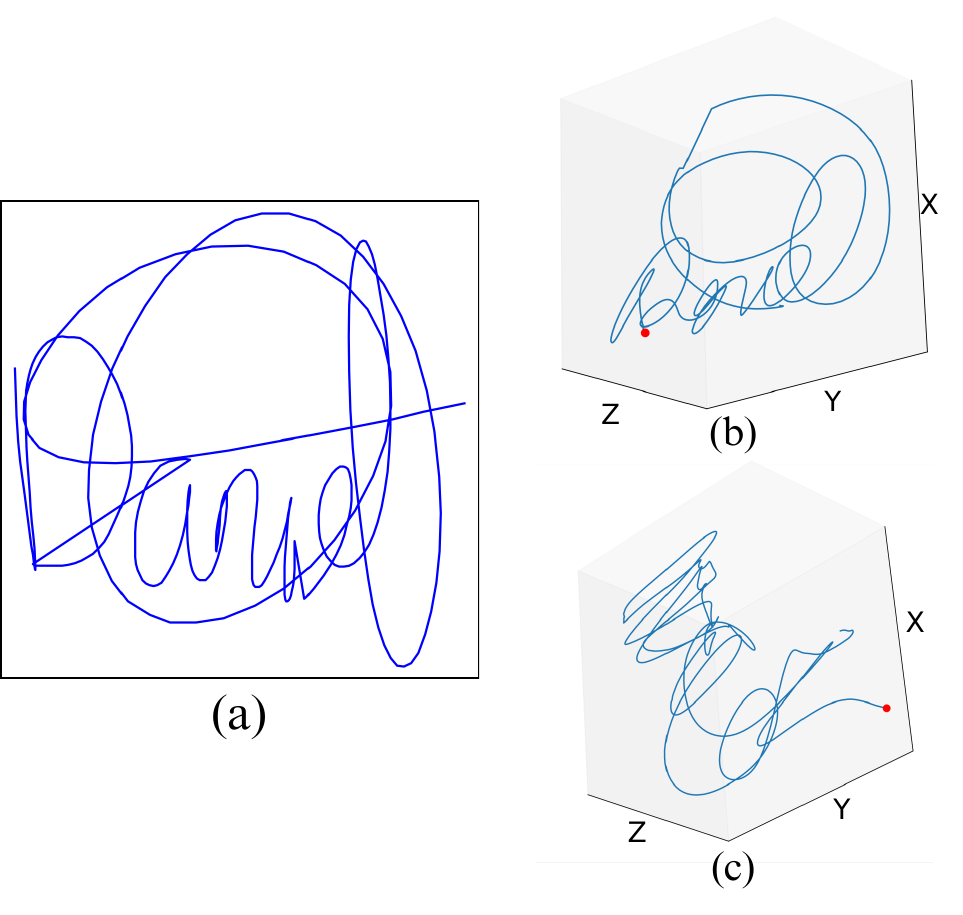}
    \caption{Two in-air signatures from the same user reconstructed using identical processing parameters. (a) shows the reference handwritten signature and (b) its corresponding 3D reconstruction. (c) presents another reconstruction attempt.}

    \label{fig:firmasdani}
\end{figure}

Although trajectory reconstruction is feasible, it is far from straightforward. Figure \ref{fig:firmasdani} illustrates two in-air signatures from the same user, reconstructed using identical processing parameters. Although the first one (Fig. \ref{fig:firmasdani}b) closely resembles the shape of the handwritten reference (Fig. \ref{fig:firmasdani}a), the other one (Fig. \ref{fig:firmasdani}c) yields a distorted and ambiguous result. This example illustrates the inherent difficulty and variability of the reconstruction task.

\vspace{1em}

\subsection{Privacy Considerations}
\label{sec:privacy}
The ability to reconstruct in-air signatures from inertial data introduces new challenges to long-standing assumptions about the privacy properties of behavioral biometrics. Conventionally, in-air gestures have been considered to be inherently ephemeral and traceless. This has led to the assertion that their privacy risk is comparatively low, in contrast to physiological traits such as 2D signature, fingerprint \cite{2017_PR_multiBtpHE_marta}, or face biometrics \cite{pietro25cancelable}. This notion has been the foundation for numerous studies and applications that treat motion-based authentication as privacy-friendly by design~\cite{guo_traceable}. However, our findings challenge this assumption by demonstrating that realistic 3D trajectories can be reconstructed from inertial signals alone.

Furthermore, the ability to transform inertial motion data into a recognizable visual signature introduces a novel form of information leakage. Despite the fact that the original signal is captured as raw sensor readings, its reconstruction can reveal content that closely resembles a traditional handwritten signature. This transformation raises significant questions about the legal treatment of such data. From a data protection point of view, these results suggest that inertial biometric data must be handled with the same care as traditionally sensitive modalities. This includes considering encryption techniques \cite{campisi10cancelable,marta17privacy} and the implementation of restrictions on access to raw sensor traces \cite{tolo23attacks-sign}.

\section{Conclusion}
\label{sec:conclusion}
This work introduces a large-scale, publicly available dataset for in-air signature verification, captured under real-world conditions using heterogeneous smartphone hardware. The dataset includes both genuine and skilled forgery samples, enabling a rigorous evaluation of biometric systems in realistic attack scenarios. We present baseline results employing classical and deep learning methods across various sensor modalities. In the random impostor setting, the lowest EER was 4.7\%, obtained using DTW with the linear accelerometer. In skilled impostor scenarios, the best performance was achieved using a fusion of all inertial signals, reaching an EER of 2.3\%. These results confirm the feasibility of the task and set realistic performance expectations in unconstrained settings. Additionally, we demonstrate for the first time the feasibility of reconstructing 3D in-air signature trajectories from inertial data alone, using handwritten references as ground truth. This challenges the long-held assumption that such signals are inherently traceless, raising important implications for privacy. Overall, our findings highlight the potential of inertial biometrics, while emphasizing the urgent need for privacy-aware design as researchers continue to expand the interpretability of sensor-based data.

\section*{Acknowledgements}
\label{sec:acks}
{\small
This study is supported by INTER-ACTION (PID2021-126521OBI00 MICINN/FEDER), HumanCAIC (TED2021-131787B-I00 MICINN), M2RAI (PID2024-160053OB-I00 MICIU/FEDER) and Cátedra ENIA UAM-Veridas en IA Responsable (NextGenerationEU PRTR TSI-100927-2023-2).}

{\small
\bibliographystyle{ieee}
\bibliography{egbib}
}

\end{document}